\documentclass[useAMS,usenatbib]{mn2e}
\usepackage{rotating}
\usepackage{journals}

\def\approxgt{\ifmmode \rlap{$>$}{}_{{}_{{}_{\textstyle\sim}}} \else%
$\rlap{$>$}{}_{{}_{{}_{\textstyle\sim}}}$\fi} 
\def\approxlt{\ifmmode \rlap{$<$}{}_{{}_{{}_{\textstyle\sim}}} \else%
$\rlap{$<$}{}_{{}_{{}_{\textstyle\sim}}}$\fi}

\LARGE \normalsize \title[2S~0921--630]{The radial velocity of the companion
star in the low--mass X--ray binary 2S~0921--630: limits on the mass
of the compact object$\dagger$}

\author[Jonker et al.]  {P.G.~Jonker$^{1}$\thanks{email :
pjonker@cfa.harvard.edu; Chandra Fellow\newline $\dagger$ Based on
observations made with ESO Telescopes at the Paranal Observatories under
programme ID 72.D--0028}, D.~Steeghs$^1$, G.~Nelemans$^2$, M.~van der
Klis$^3$\\ $^1$Harvard--Smithsonian Center for Astrophysics, 60 Garden Street,
Cambridge, MA~02138, Massachusetts, U.S.A.\\ $^2$Institute of Astronomy,
Madingley Road, CB3 0HA, Cambridge, UK\\ $^3$Astronomical Institute ``Anton
Pannekoek'', University of Amsterdam, Kruislaan 403, 1098 SJ Amsterdam, The
Netherlands\\}

\begin{document}

\maketitle

\begin{abstract} \noindent  In this Paper we report on optical spectroscopic
observations of the low-mass X-ray binary 2S~0921--630 obtained with
the Very Large Telescope. We found sinusoidal radial velocity
variations of the companion star with a semi--amplitude of
99.1$\pm$3.1 km s$^{-1}$ modulated on a period of 9.006$\pm$0.007
days, consistent with the orbital period found before for this source,
and a systemic velocity of 44.4$\pm$2.4 km s$^{-1}$.  Due to X--ray
irradiation the centre--of--light measured by the absorption lines
from the companion star is likely shifted with respect to the
centre--of--mass. We try to correct for this using the so--called
K--correction.  Conservatively applying the maximum correction
possible and using the previously measured rotational velocity of the
companion star, we find a lower limit to the mass of the compact
object in 2S~0921--630 of $M_{X} \sin^3 i > 1.90\pm0.25\,M_\odot$ (1
$\sigma$ errors). The inclination in this system is well constrained
since partial eclipses have been observed in X-rays and optical
bands. For inclinations between 60$^\circ < i< 90^\circ$ we find
$1.90\pm0.25< M_{X} < 2.9\pm0.4\,M_\odot$. However, using this maximum
K--correction we find that the ratio between the mass of the companion
star and that of the compact object, q, is $1.32\pm0.37$ implying
super--Eddington mass transfer rates; however, evidence for that has
not been found in 2S~0921--630.  We conclude that the compact object
in 2S~0921--630 is either a (massive) neutron star or a
low--mass~black~hole.
\end{abstract}


\section{Introduction} \label{intro} The equation of state (EoS) of neutron
star matter is intimately related to the physics of the strong interactions
between fundamental particles and therefore it is of great relevance in
high--energy and particle physics. Theories on the EoS of neutron--star matter
at supra--nuclear densities provide a firm upper limit on the mass for each
EoS. The stiffer the EoS, the higher the mass limit. Neutron stars with masses
well above $1.4\,M_\odot$ cannot exist for so--called soft equations of state,
in which matter at high densities is relatively compressible (e.g., due to
meson condensation or a transition between the hadron and quark--gluon phases;
cf.~\citealt{2001ApJ...550..426L}). Therefore, measuring a high mass for even
one neutron star would imply the firm rejection of many proposed soft EoSs
(e.g.~see discussion by \citealt{1995xrb..book...58V}).  Neutron star masses of
more than $\sim 3\,M_{\sun}$ are excluded assuming (among other things) that
general relativity is the correct theory of gravity and that the velocity of
sound is less than the velocity of light (\citealt{1973ApJ...179..277N}). 
Therefore, a lower limit on the mass of the compact object of 3 $M_{\odot}$ or
more implies a black hole compact object.

The most accurately measured neutron star masses we have are from
precise radio timing measurements of double neutron--star
binaries. These empirical masses cluster around $1.35\,M_\odot$
(\citealt{1999ApJ...512..288T}). This is close to the mass of 1.32
$M_{\odot}$ predicted from theoretical model calculations of Type Ib
supernovae (\citealt{1996ApJ...457..834T}). However, these double
neutron star binaries are formed in high--mass X--ray binaries and due
to the short life of the massive companion only a limited amount of
matter can be accreted. In contrast, in low--mass X--ray binaries
(LMXBs), the rapid rotation and low magnetic field of the neutron
stars are considered evidence for much larger amounts of accreted
matter (see \citealt{bhatta1995} for a review).  Indeed, from binary
evolution models, LMXBs could accrete up to 0.7\,$M_\odot$
(e.g.~\citealt{1995A&A...297L..41V}). If so, the masses of their
neutron stars could be $>\!2\,M_\odot$. Most of the constraints from
timing observations of the successors of these LMXB primaries, the
millisecond radio pulsars, are weak (\citealt{1999ApJ...512..288T}),
as the almost purely circular orbits of these systems often prevents
one from measuring most of the post--Newtonian parameters that
underlie the precise masses for the double--neutron star
binaries. With masses of $1.57^{+0.25}_{-0.2}\,M_\odot$ and
2.2$\pm0.4\,M_\odot$, the msec pulsars PSR~B1855$+$09 and
PSR~J0751+1807 are exceptions to this (\citealt{2003rapu.conf...75N};
\citealt{2004aspin.nice},~95\%~conf.).

Dynamical mass estimates or dynamically determined limits on the mass of the
neutron star in LMXBs are available for the two transient sources Cen~X--4 and
XTE~J2123--058, the Z--source Cyg~X--2, and the pulsar 2S~1822--371 (Cen~X--4;
$M_{NS}=1.3\pm$0.6 $M_\odot$ \citealt{1993MNRAS.265..655S}; XTE~J2123--058;
$M_{NS}=1.55\pm$0.31 $M_\odot$ \citealt{2002MNRAS.329...29C}; Cyg~X--2;
$M_{NS}=1.78\pm$0.23 $M_\odot$ \citealt{1998ApJ...493L..39C};
\citealt{1999MNRAS.305..132O}; 2S~1822--371; $M_{NS}>1.14\pm$0.06 $M_\odot$;
\citealt{2003MNRAS.339..663J}; \citealt{2003ApJ...590.1041C}). The problem is
that many of the companion stars are small, late type stars which at a distance
of more than a few kpc are too faint to observe, especially since the
integration times are limited to typically 1/20th of the orbital period in
order to avoid Doppler smearing of the spectral lines due to the binary motion.
Furthermore, the binary inclination is often poorly constrained and effects of
irradiation shift the centre-of-light with respect to the centre-of-mass. 

Some of the aforementioned problems will be less severe when observing long
orbital period systems for which the inclination is well constrained from the
fact that eclipses are observed: 2S~0921--630/V395~Car is such a high
inclination LMXB with a long orbital period (P$_{orb}\approx 9.0$ days;
e.g.~\citealt {1982ApJ...256..605C}, \citealt{1983MNRAS.205..403B}, and
\citealt{1987MNRAS.226..423M}). The inclination of 2S~0921--630 is relatively
well constrained and must be high since partial eclipses of the compact object
and accretion disc have been observed in both X--rays and optical wave bands
(e.g.~\citealt{1981SSRv...30..279B}; \citealt{1981A&A....94L...3C};
\citealt{1987MNRAS.226..423M}). 2S~0921--630 is thought to be a halo object in
orbit with a K0--1 III companion star; absorption lines of the companion have
been detected  (\citealt{1983MNRAS.205..403B}; \citealt{1999A&A...344..101S}). 
The corresponding radial velocity curve would provide a limit to the mass of
the companion star. In this Paper we report on spectroscopic observations of
2S~0921--630/V395~Car. 

\section{Observations and analysis} 

We observed 2S~0921--630/V395~Car with the FORS2 spectrograph mounted
on the Yepun Very Large Telescope (VLT).  In the period Dec 26, 2003
-- Mar 13, 2004, 22 spectra using the 1200R+93 and 22 spectra using
the 1028z+29 grating have been obtained in Service Mode. The exposure
time was 1300~s for each observation and a slit--width of 0.4" was
used on each occasion. The dispersion was 0.75 \AA~per pixel for the
spectra obtained with the 1200R+93 grating and 0.86 \AA~per pixel for
the spectra obtained with the 1028z+29 grating. With a slit--width of
0.4" the two--pixel resolution is approximately 65 km s$^{-1}$ at 6500
\AA.  After each observing night Helium--Neon--Argon lamp wavelength
calibration spectra were obtained. The seeing varied between 0.5'' and
1.4'' and the spectra were obtained when the source had an airmass of
$\sim1.2-1.4$, except on one occasion on Jan 20, 2004 when the airmass
was nearly 2.

The spectra were bias subtracted, flatfield corrected, optimally
extracted and wavelength calibrated using the
\textsc{iraf}\footnote{\textsc {iraf} is distributed by the National
Optical Astronomy Observatories } reduction package. The rms scatter
in the wavelength calibration was $\approx$0.03\AA. We further applied
a small correction to the wavelength solution by aligning sky lines
observed in each of the spectra. Due to variations in seeing,
resulting in variable slit losses, the signal--to--noise ratio varied
between the spectra from $\approx$80--240 over the wavelength range
5920--6520 \AA. Next, we exported the extracted spectra to the data
analysis package \textsc {molly}.

\begin{figure} \includegraphics[width=8cm]{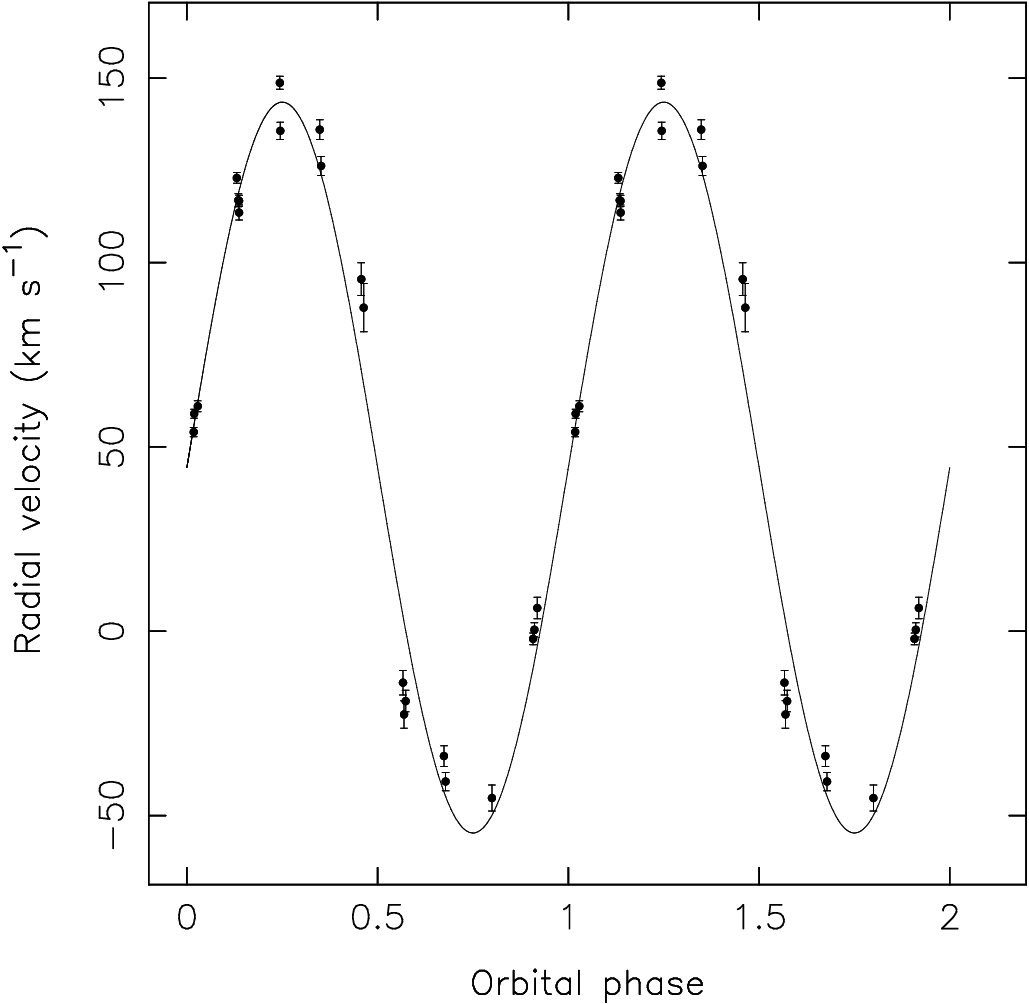} \caption{Radial
velocity of the companion star in 2S~0921--630. Two cycles  have been plotted
for clarity. Phase zero is superior conjunction of the neutron star.
Overplotted is the best-fit sinusoidal radial velocity curve. The data points
are shown with their formal statistical errors only.} \label{radvel} \end{figure}

In \textsc {molly} the observation times were first corrected to the
Heliocentric Julian Date time frame (using UTC times). Next, we
normalised and rebinned the spectra to a uniform velocity scale
removing the Earth's velocity.  Below, we first discuss results
obtained using the 1200R+93 grating spectra. We cross--correlated the
1200R+93 grating spectra with spectra of template stars rebinned to
the same velocity scale. Nine template star spectra with spectral
types ranging from G5--K7 had been obtained with the Keck telescope at
similar resolution. We fitted a sinusoid to the measured velocities as
a function of time. The fit-parameters are: the phase, the
semi--amplitude, and the period of the sinusoid and the systemic
velocity. However, since the errors on the velocity represent the
statistical error bars only, they do not include the systematic
effects which we found to be important (we will come back to this in
the Discussion). Therefore, the reduced $\chi^2$ of the fit was much
larger than 1. In order to estimate realistic errors on the fit
parameters, we increased the size of the error bars such that the
reduced $\chi^2$ was close to 1. We found a best--fit period of
9.006$\pm0.007$ days, consistent with the orbital period of $\approx$9
days found before (here and below we give 1 $\sigma$ error bars). From
this we derive an orbital ephemeris of HJD = 2453000.49(8) +
N$\times9^d.006(7)$ (UTC) where phase zero is defined as superior
conjunction of the neutron star and the number in between brackets
denotes the uncertainty in the last digit. The derived ephemeris does
not significantly depend on the spectral type of the used template
star and is consistent within the errors with the ephemeris given by
\citet{1987MNRAS.226..423M}. We determined a radial velocity semi--amplitude of
99.1$\pm$3.1 km s$^{-1}$ using the K1V template star HD~124106 (see
Figure~\ref{radvel}). The radial velocity semi-amplitude does not significantly
depend on the spectral type of the template star. I.e.~the  amplitudes we found
varied by 2 km s$^{-1}$ between the minimum and maximum amplitude for template
spectral types in the range of G5--K7. We found a systemic velocity of
44.4$\pm$2.4 km s$^{-1}$. 

\begin{figure*}
\includegraphics[width=4.5cm,height=8.5cm,angle=-90]{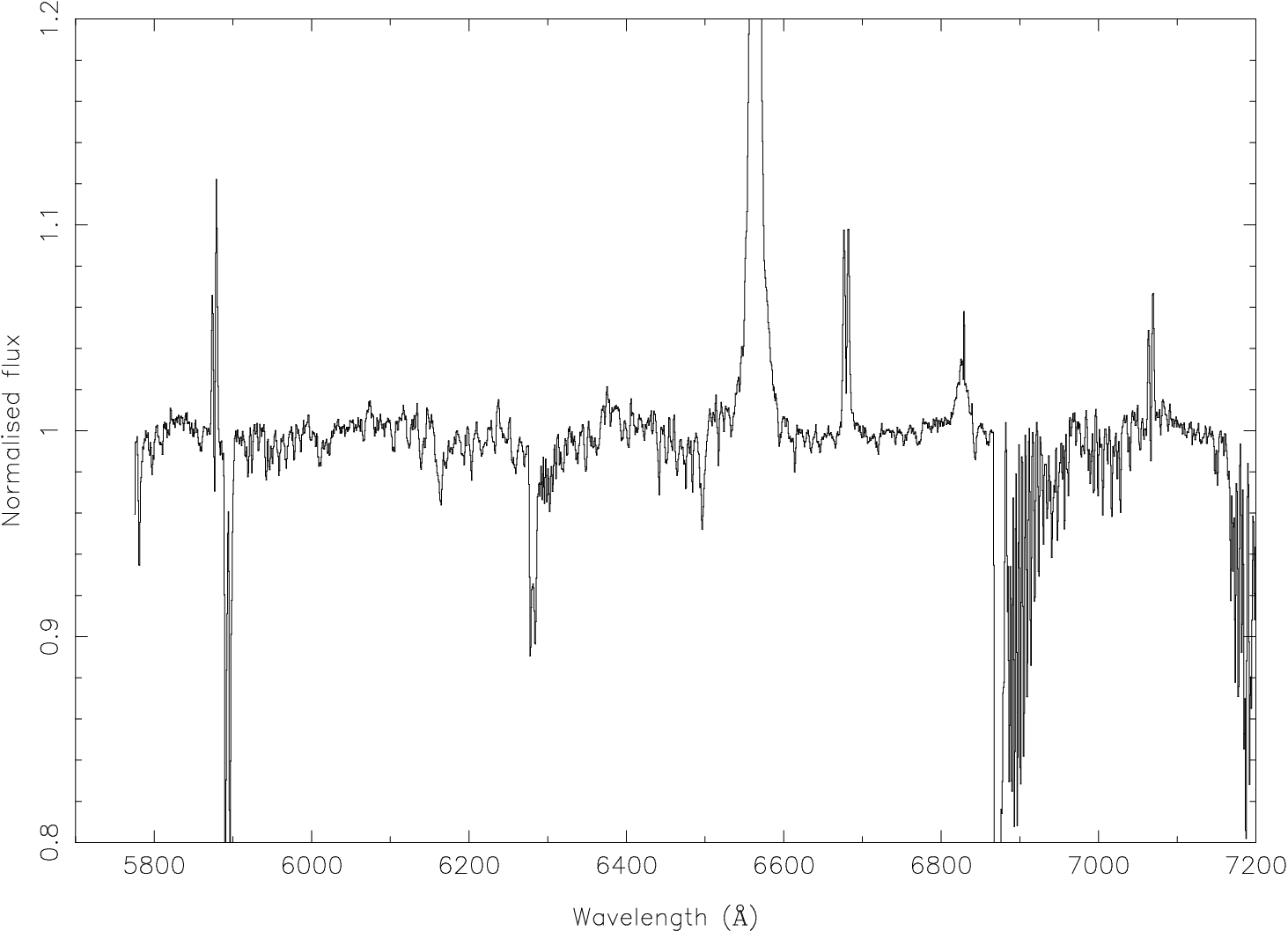} \quad
\includegraphics[width=4.5cm,height=8.5cm,angle=-90]{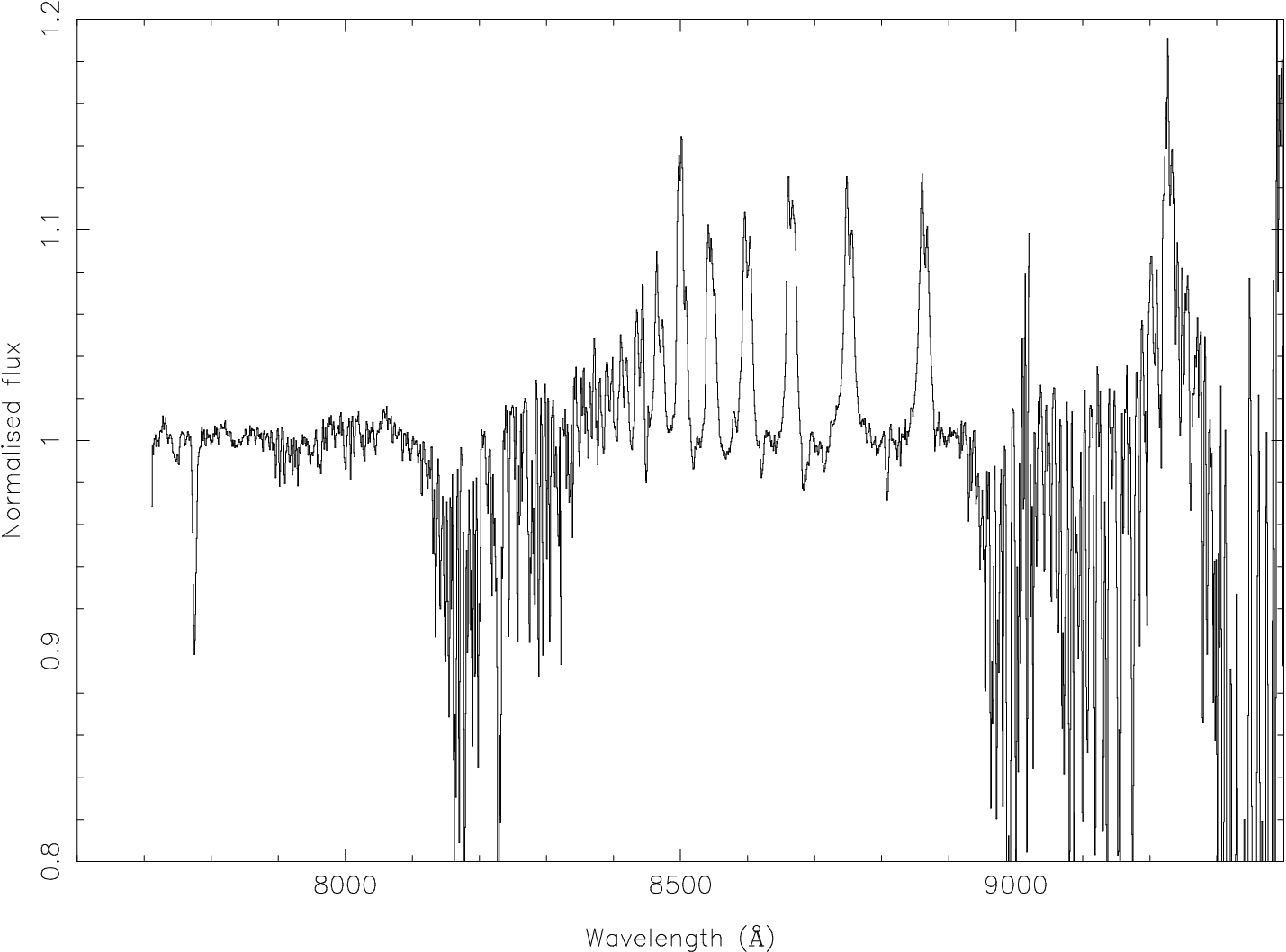} \quad
\includegraphics[height=9cm]{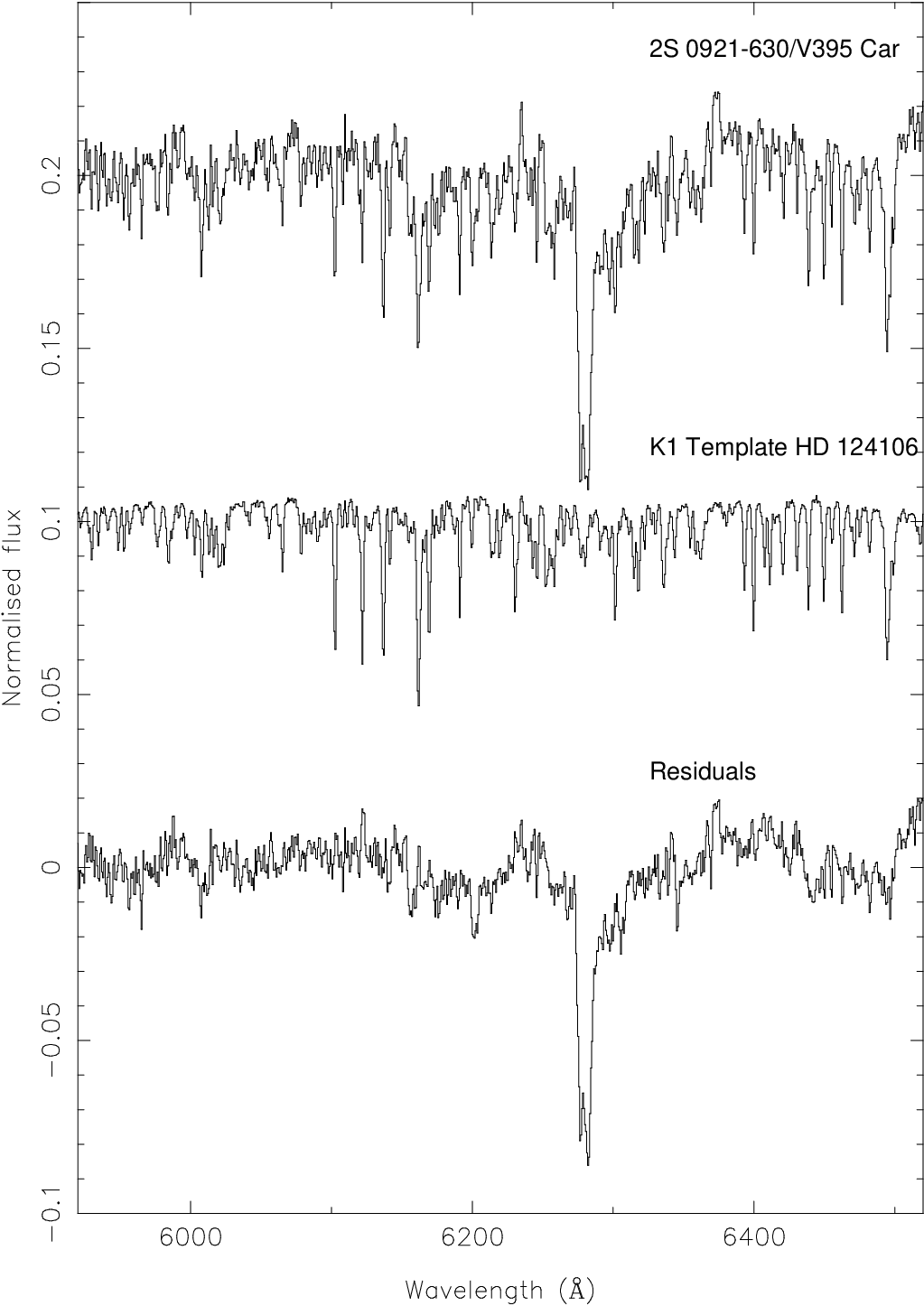} \quad
\includegraphics[height=9cm]{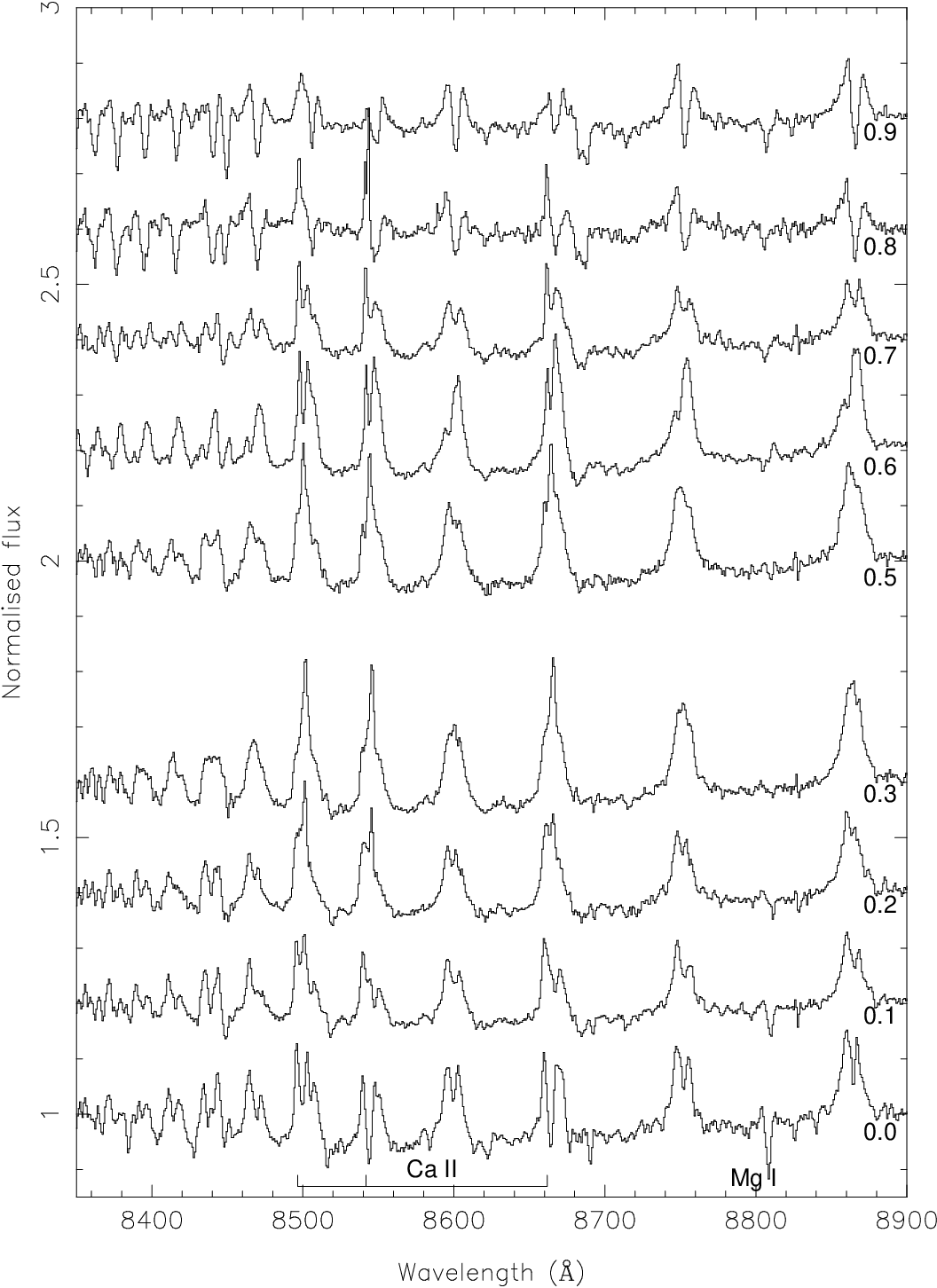} \caption{
{\it Top left panel: } The spectrum after averaging all the spectra
obtained with the 1200R+93 grating. {\it Top right panel: } Same as
the {\it top left panel} but using spectra obtained with the 1028z+29
grating. {\it Bottom left panel:} From top to bottom: The average
spectrum of 2S~0921--630 in the range 5920--6520 \AA in the frame of
the companion star. The K1 template star (HD~124106) spectrum veiled
by an accretion disc contribution of 80 per cent.  The residual
spectrum after subtraction of the veiled template star spectrum. For
clarity, the spectra have been normalised and shifted vertically.
{\it Bottom right panel:} Average spectra as a function of phase, the
number on the right indicates the centre of the orbital phase bin
(there are no spectra in the phase bin from 0.35--0.45 obtained with
the 1028z+29 grating hence only 9 spectra are plotted). The CaII and
MgI lines from the companion star have been identified below the
average spectrum at phase 0.0.}
\label{optsub}
\end{figure*}

After correcting each individual spectrum for the observed sinusoidal velocity
shift appropriate for the orbital phase we created an average spectrum in the
frame of the donor star for the data of 2S~0921--630. One can in principle
derive the rotational broadening from an optimal subtraction (see for the
description of this technique \citealt{1994MNRAS.266..137M}). However, our data
is not well suited for this since the dispersion at which both the template
star and the source were observed is relatively low ($\approx30$ km s$^{-1}$
per pixel in the 5920--6520 \AA~spectral range for the template star, this
spectrum was resampled to the dispersion of $\approx34$ km s$^{-1}$ per pixel
in the 5920--6520 \AA~spectral range at which the source was observed). Since
the template stars were not obtained with the same telescope and instrumental
set--up, the instrumental profile will be different from that of the object
spectra; together with the low resolution this means that it is not possible to
reliably measure the rotational broadening when it is of the order of the
spectral resolution. Earlier results on the rotational broadening of 64$\pm$9
km s$^{-1}$ measured by \citet{1999A&A...344..101S} showed that this is likely
the case. \citet{1999A&A...344..101S} found a best--fit spectral type of K0 for
the companion star; we found that the residuals were smallest when we used the
K1 template star HD~124106 (irrespective of broadening). We further found that
on average 20 per cent of the light comes from the companion star in the 
5920--6520 \AA~range. In Figure~\ref{optsub} ({\it Left panel}) we plot the
average spectrum in this range, the broadened template star veiled by a source
of constant light contributing 80 per cent of the light, and the residuals
after subtraction the veiled template star spectrum. We used the spectral range
from 5920--6520 \AA~for the optimal subtraction, however, the range in
wavelengths between 6250--6350 \AA~was masked since it contains a strong
interstellar absorption feature.

\begin{figure} \includegraphics[width=8cm]{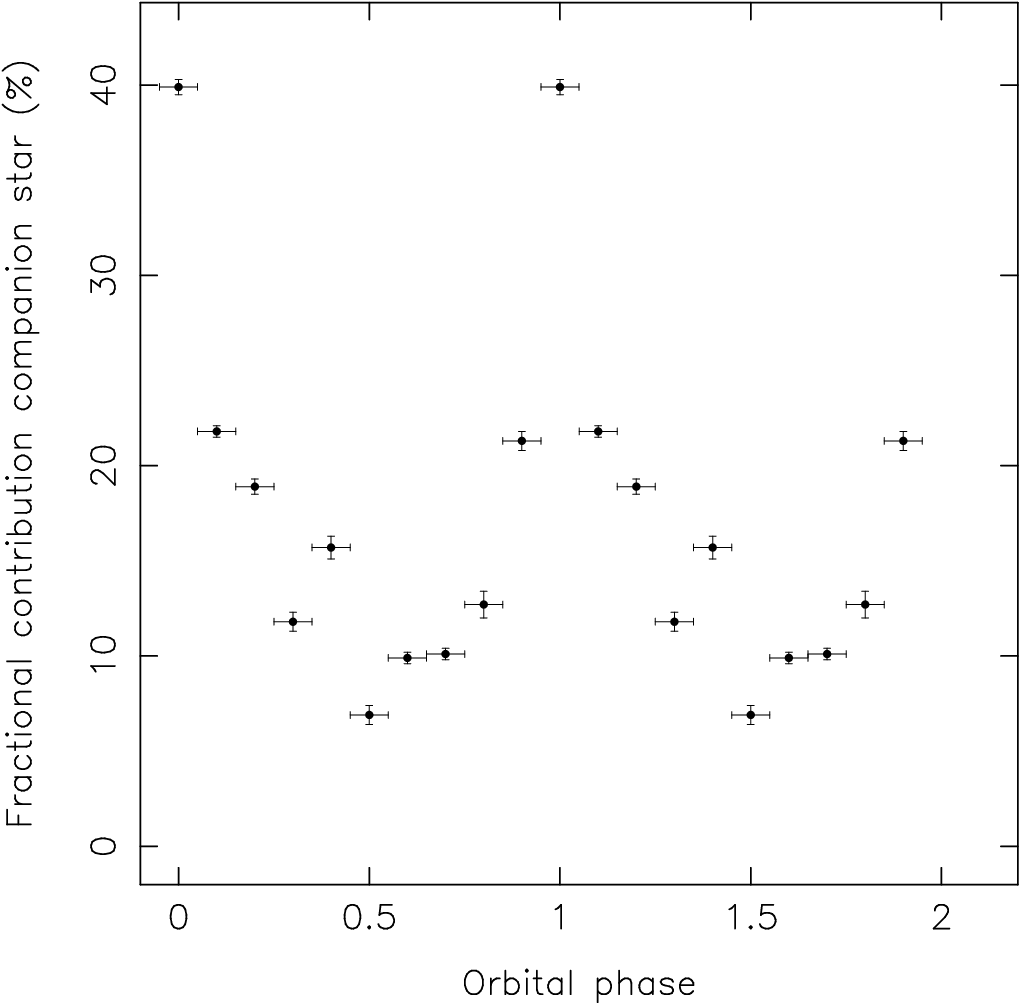} \caption{The
fractional contribution of the K1 companion star to the total light in the
5920--6520 \AA\,wavelength range as a function of orbital phase. Two cycles
have been plotted for clarity. Phase zero is superior conjunction of the
neutron star.} \label{compfrac} \end{figure}

We further examined whether the contribution of the K1 companion star to the
total amount of light in the 5920--6520 \AA~range varies as a function of the
binary orbital phase. We averaged spectra obtained at the same phase using
phase bins of width 0.1. We optimally subtracted the template spectrum of the
K1 star from the average spectra in each phase bin. The measured fractional
contribution of the companion star to the total amount of light in the
5920--6520 \AA~range is plotted in Figure~\ref{compfrac}. The large difference
in fractional contribution from the K--star between phase 0 and phase 0.1 is
difficult to explain. 

Besides the stellar absorption lines apparent in the wavelength range
5920--6520 \AA~shown in the top spectrum of the {\it Bottom left
panel} in Figure~\ref{optsub}, additional stellar absorption lines
were present in the range 6520--6800 \AA (see the {\it Top left panel}
in Figure~\ref{optsub}). Furthermore, several emission lines are
present in the spectrum. Unfortunately, our observations lack the
resolution to resolve these accretion disc lines and hence only one
broad emission peak was observed whereas
e.g.~\citet{1999A&A...344..101S} observed that the H$\alpha$ and He~I
emission lines at 6562.76 \AA~and 6678.15 \AA, respectively have the
double peaked profile typically observed in accretion disc spectra. We
detected a broad emission peak at $\approx$6820 \AA~which we could not
positively identify. We checked whether CCD defects were present at
the position relevant for this wavelength in the flatfield images or
in the individual images but none were found.

So far, we have only discussed spectra obtained with the 1200R+93
grating. In Figure~\ref{optsub} ({\it Bottom right panel}) we show the
phase folded average spectra of the 1028z+29 grating in the range
between 8300--8900 \AA. Clear donor star features such as the Ca~II
triplet at 8498 \AA, 8542 \AA, and 8662 \AA, and the Mg~I line at
8806.75 \AA~can be seen at orbital phases where the back,
non--irradiated side, of the companion star is observed. Furthermore,
superimposed on the Paschen disc emission lines, Paschen absorption
lines can be seen. Unfortunately, the combined effect of two
unresolved broad accretion disc emission lines, a Paschen absorption
line and, at some wavelengths, and at certain orbital phases, a Ca~II
absorption line is too difficult to disentangle with the limited
resolution data in hand. Therefore, we did not use these spectra for
our radial velocity study. Besides the features visible in the {\it
bottom right panel} of Figure~\ref{optsub} we identified two other
lines in the 1028z+29 grating spectra; a weak Fe~I absorption line
complex near 7750 \AA~and a Paschen emission line at 9229~\AA (see the
{\it top right panel} in Figure~\ref{optsub}).

\section{Discussion}

We determined the radial velocity curve of the companion star in
2S~0921--630/V395~Car using VLT/FORS2 spectroscopic observations. A sinusoidal
fit to the radial velocity measurements with a semi--amplitude, K$_2$, of
99.1$\pm$3.1 km s$^{-1}$ and with a systemic velocity, $\gamma=$44.4$\pm$2.4 km
s$^{-1}$ represents the data well. In these error estimates we artificially
increased the error bars on the individual velocity measurements by a factor 4
in order to obtain a reduced $\chi^2$ of the fit close to 1. Before we discuss
the implications of the K--velocity, we investigate possible reasons for the
large reduced $\chi^2$ obtained when using the formal statistical errors only.

Near quadrature at phase 0.25 the velocity measurements obtained on different
nights differ well beyond the statistical errors.  From Figure~\ref{compfrac}
we can see that the companion star contribution to the total amount of light in
the 5920--6520 \AA~range varies as a function of orbital phase. The phasing of
this variation suggests that the centre of light emitted by the K--star is
likely to be shifted with respect to its centre of mass. An explanation is that
the inner side of the companion star is heated by the (X--ray) irradiation
coming from (near) the compact object, reducing the equivalent width of the
K--star stellar lines in the range 5920--6520 \AA~(see also
e.g.~\citealt{1988ApJ...324..411W}). This manifests itself in several
ways:\newline {\it (i)} a lower fractional contribution to the total amount of
light from the K star at e.g.~phase 0.5 when using the optimal subtraction
technique (e.g.~Figure~\ref{compfrac})\newline {\it (ii)} an increase in
velocity (especially near quadrature) with respect to the velocity associated
with the centre of mass of the companion star, since the line profiles will, as
a result of the stellar rotation, be preferentially shifted to either a larger
red- (phase 0.25) or blue-shift (phase 0.75). Hence, variations in the amount
of (X--ray) irradiation between the observations could explain the variations
in velocities near quadrature as observed in Figure~\ref{radvel}. Since the
observations were obtained with at least one day in between, with a baseline of
several months, it seems likely that the irradiation changed between the
observations.

Furthermore, the velocity measurements around phase 0.5 do not fall on the
best--fit sinusoid. Those at phases 0.4--0.5 lay above the best--fit sinusoid,
whereas those at phases 0.5--0.6 fall below the curve. This could well be
explained by the Rossiter effect (cf.~\citealt{2001icbs.book.....H}). Such
deviations occur if the companion star is partially eclipsed by the accretion
disc. Such a partial eclipse of the companion star can also help explain the
low fraction of light coming from the companion star at phase 0.5 (see
Figure~\ref{compfrac}). The radial velocity measurements at phases just prior
to phase 0.5, the partial eclipse of the companion star, are biased to higher,
red shifted velocities since the part of the companion star that is eclipsed at
those phases is mainly rotating towards the observer. Hence, the rotational
broadening of the line will be biased towards the red. The reverse holds for
phases just after the partial eclipse. 

For the rotational broadening of the stellar lines we take $v \sin
i=64\pm9$ km s$^{-1}$ as found by \citet{1999A&A...344..101S}.  For a
Roche lobe filling companion star the following relation holds:
$\frac{{\rm v\,sin}i}{{\rm K}_2}=0.46[(1+q)^2q]^\frac{1}{3}$
(e.g.~\citealt{1988ApJ...324..411W}; q is defined here as M$_2$
divided by the mass of the compact object, M$_X$). Before we can
determine q, and hence with K$_2$ the mass of the compact object, we
have to take effects of the non--uniform absorption distribution on
the stellar surface caused by irradiation into account. To this end,
\citet{1988ApJ...324..411W} developed a so--called K--correction. Following the
procedure laid out in \citet{1988ApJ...324..411W}, ${\rm K_{2,corr}=K_{obs} -
\Delta K}$.  $K_{2,corr}$ is the corrected observed ($K_{obs}$) radial velocity
semi--amplitude and $\Delta K = f R_2 K_2/a_2 = f v \sin i$. Here
$a_2$ is the distance between the centre-of-mass of the binary and the
centre-of-mass of the companion star and $f$ is a geometrical
correction factor less than 1. In the extreme case that the hemisphere
facing the compact object does not contribute to the observed stellar
absorption lines at all and the other hemisphere has a uniform
absorption line strength, $f=\frac{4}{3\pi}$
(\citealt{1988ApJ...324..411W}). Since we still have a contribution of
the K1 star of around 7 per cent at phase 0.5 we know that this is an
overestimation of the K--correction in the case of
2S~0921--630. Hence, applying this upper limit on the K--correction,
$\Delta K \leq 26.9\pm3.8$ km s$^{-1}$, we get a firm lower limit of
K$_{2,corr}\geq 72.2\pm4.9$ km s$^{-1}$. Plugging--in the numbers
given above yields q$=$1.32$\pm0.37$; note that in deriving this we
used the equation for $R_2/a$ of \citet{paczynski1971} which is valid
for $0<q<0.8$. However, the difference with the equation of
\citet{1983ApJ...268..368E} is small for q$<$1.7 and does not change
the result significantly given the large uncertainty in $v \sin
i$. The uncertainty in q is dominated by the large error on the
rotational velocity. Thus, we find for the minimum mass of the compact
object, $M_X \sin^3 i > 1.90\pm0.25\,M_\odot$. As mentioned in the
Introduction, the inclination of 2S~0921--630 must be high.
Furthermore, we have presented evidence for a partial eclipse of the
companion star. For inclinations, $60^\circ<i<90^\circ$
(cf.~\citealt{1983MNRAS.205..403B}; \citealt{1987MNRAS.226..423M})
the mass of the compact object is
$1.90\pm0.25<M_X<2.9\pm0.4\,M_\odot$.

Using $v \sin i = 2 \pi R_2 \sin i/P$, we get for the radius of the companion
star, $R_2$, $ R_2 \sin i= 11.4\pm1.6\,R_\odot$. Under the assumption that the
companion star fills its Roche lobe, the mass of the donor star, $M_2$, depends
only on its radius and the orbital period (cf.~\citealt{paczynski1971}), which
gives $M_2 \sin^3 i= \frac{(R_2\sin i)^3}{0.234^3 \times P_{orb}^2}$ (here the
orbital period is measured in hours, and the mass and radius of the companion
star are in solar units). This gives $M_2 \sin^3 i=2.5\pm1.1\,M_\odot$. This
M$_2$ suggests an interesting evolutionary stage of the binary. M$_2$ is more
massive than standard donor stars in LMXBs. Recently, there have been a number
of evolutionary calculations (e.g.~\citealt{2000ApJ...530L..93T};
\citealt{2002ApJ...565.1107P}) showing that systems with donor masses as high
as $\sim$5 M$_\odot$~can enter a relatively long phase of mass transfer to a
neutron star after the donor has lost most of its mass. 2S~0921--630 would
nicely fit in this scenario, evolving along the track of an initial 3--4
M$_\odot$~donor. However, in this evolutionary scenario the current
time-averaged mass-transfer rate is expected to be high, certainly
super-Eddington.  This is at odds with the finding from studying X-ray lines as
measured with {\it Chandra} and XMM--{\it Newton} that the observed X-ray
luminosity of $\sim10^{36}$ erg s$^{-1}$ (for a distance of $\sim$7 kpc) is
likely close to the intrinsic source luminosity
(\citealt{2003ApJ...583..861K}). Perhaps the current mass-transfer rate is much
lower than the time-averaged transfer rate, or the donor mass and/or the
mass-ratio are overestimated.

Above, we assumed that the companion star is in co--rotation with the
binary orbit. Synchronization of the rotational period and the binary
period occurs on timescales shorter than the circularization timescale
(e.g.~\citealt{1988ApJ...324L..71T}) and no signs for orbital
eccentricity have been found (see Figure~\ref{radvel}). Furthermore,
using the companion star parameters derived above and equation 7 of
\citet{1988ApJ...324L..71T} we find that the synchronization timescale
is $\approxlt$100 yr for 2S~0921--630, justifying this assumption.  We
conclude that the compact object in 2S~0921--630 is likely to be a
massive neutron star or a low-mass black hole. If indeed the compact
object in 2S~0921--630 is a massive neutron star the mass measurement
would rule out soft equations of state.  Future high resolution
optical observations should provide a much more accurate handle on the
rotational velocity of the companion star, yielding a more accurate
constraint on the mass of the compact object.

\section*{Acknowledgments}  

\noindent Support for this work was provided by NASA through Chandra
Postdoctoral Fellowship grant number PF3--40027 awarded by the Chandra X--ray
Center, which is operated by the Smithsonian Astrophysical Observatory for NASA
under contract NAS8--39073. DS acknowledges a Smithsonian Astrophysical
Observatory Clay Fellowship. GN is supported by PPARC. The use of the spectral
analysis software package \textsc {molly} written by Tom Marsh is acknowledged.

\end{document}